# Identifying Crisis Response Communities in Online Social Networks for Compound Disasters: The Case of Hurricane Laura and Covid-19

**Khondhaker Al Momin**
Graduate Research Assistant
School of Civil Engineering & Environmental Science
University of Oklahoma
202 W. Boyd St., Norman, OK 73019-1024
Email: momin@ou.edu

**H M Imran Kays**
Graduate Research Assistant
School of Civil Engineering & Environmental Science
University of Oklahoma
202 W. Boyd St., Norman, OK 73019-1024
Email: kays@ou.edu

**Arif Mohaimin Sadri, Ph.D.**
Assistant Professor
School of Civil Engineering & Environmental Science
University of Oklahoma
202 W. Boyd St., Norman, OK 73019-1024
Email: sadri@ou.edu
*(Corresponding Author)*

## ABSTRACT

Online social networks allow different agencies and the public to interact and share the underlying risks and protective actions during major disasters. This study revealed such crisis communication patterns during hurricane Laura compounded by the COVID-19 pandemic. Laura was one of the strongest (Category 4) hurricanes on record to make landfall in Cameron, Louisiana. Using the Application Programming Interface (API), this study utilizes large-scale social media data obtained from Twitter through the recently released academic track that provides complete and unbiased observations. The data captured publicly available tweets shared by active Twitter users from the vulnerable areas threatened by Laura. Online social networks were based on Twitter's user influence feature (i.e., mentions or tags) that allows notifying other users while posting a tweet. Using network science theories and advanced community detection algorithms, the study split these networks into twenty-one components of various size, the largest of which contained eight well-defined communities. Several natural language processing techniques (i.e., word clouds, bigrams, topic modeling) were applied to the tweets shared by the users in these communities to observe their risk-taking or risk-averse behavior during a major compounding crisis. Social media accounts of local news media, radio, universities, and popular sports pages were among those who involved heavily and interacted closely with local residents. In contrast, emergency management and planning units in the area engaged less with the public. The findings of this study provide novel insights into the design of efficient social media communication guidelines to respond better in future disasters.

**INTRODUCTION**

Natural, man-made, and technological disasters have major adverse consequences for society and the economy. Affected communities must be well-informed about the potential distress they may encounter during a disaster in order to take appropriate preparation mentally and logistically. Emergency management authorities must be aware of the needs and concerns of affected people to respond effectively throughout a crisis (*1*). Understanding the communication pattern, perspectives, thoughts, and needs of the affected people, as well as proper information dissemination, is pivotal during any disaster or emergency situation.

On August 27, 2020, hurricane Laura, one of the most powerful and deadliest "*category 4*" hurricane made landfall with peak intensity in Cameron, Louisiana (*2*) in the middle of the Covid-19 pandemic. An estimated 20 million people were in the path of the storm, with 1.5 million in Texas and Louisiana being ordered to evacuate (*3*). Hurricane Laura spawned a 15-feet storm surge that drenched some regions with 10 inches of rain (ref. **Figure 1**) and spawned four tornadoes (*4*). Lesser Antilles, Greater Antilles, Bahamas, Gulf Coast of the United States (U.S.), the Midwestern and Eastern U.S. have all been heavily affected by hurricane Laura. It largely impacted Louisiana, and a total of 14 persons were killed (10 deaths in Louisiana and 4 in Texas) in the U.S., 31 in Haiti, and 3 in the Dominican Republic (*5-7*). More than 750,000 people experienced a significant power outage due to hurricane Laura (*8*), while many refineries in the U.S. were shut down for several weeks pending repairs and power restoration in the aftermath of hurricane Laura (*9*). Moreover, this "category 4" hurricane struck the U.S. during a pandemic, when people were instructed to keep their distance. Such crisis further escalates when there is a communication gap between vulnerable communities and emergency management agencies (*10*).

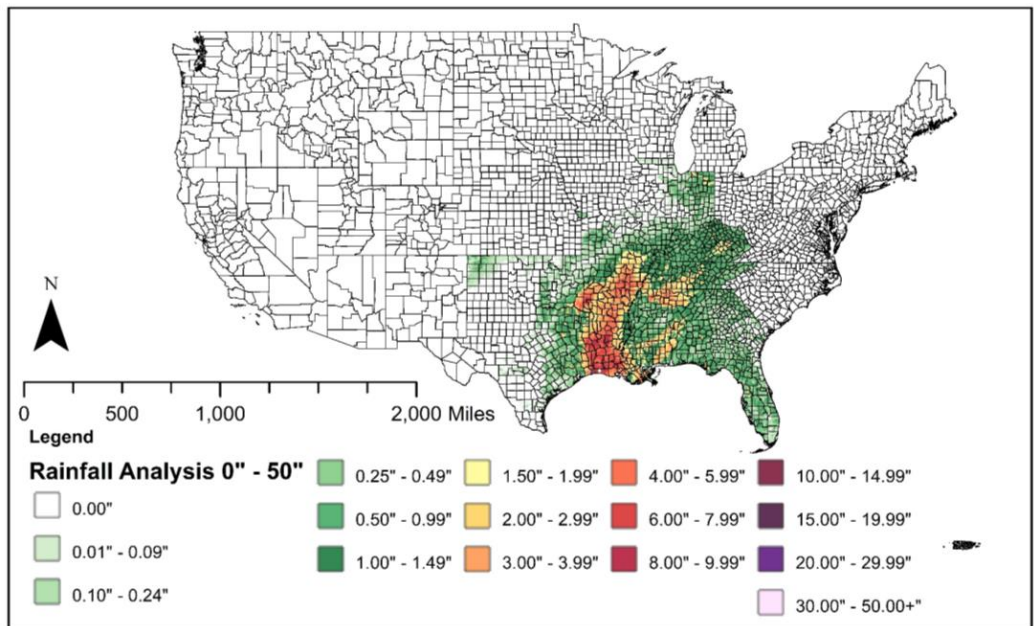

**Figure 1: Heavy Rainfall due to Hurricane Laura**

According to the U.S. Census Bureau, 84 percent of U.S. households own a cell phone, and 78 percent own a desktop or laptop computer (*11*). During a disaster, people can communicate in person or via mobile phone. However, by using online social media platforms (SMPs), they can reach and connect with more people in a much shorter period of time, which is why the use of online social media during disasters has recently increased (*12*). Many studies have shown that weather and situational awareness information is widely disseminated through SMPs (i.e., Facebook, Twitter, Reddit, Instagram, etc.) (*13-17*). These SMPs allow users to share ideas, thoughts, and information through virtual networks in a timely manner (*18*). Users readily share real-time information on social media about a variety of topics such as disasters, potential crises, weather updates, and traffic updates, among others. According to a recent report (*19*), there are approximately 450 million monthly active Twitter users (*20*), and among them, more than 80.9 million users are from the U.S. The majority of the emergency management and law enforcement agencies in the U.S. have a Twitter account where they share important and relevant information (*21*). People follow different organizations as well as different celebrities on social media, and a lot of users are directly influenced by their followees on SMPs (*22*).

The Covid-19 outbreak saw more than 29.7% more users spending 1-2 hours per day on social media, with 20.5% using SMPs 30 minutes to an hour more (*18,23*), and even during a power outage, people can still use different social media to keep communication with their friends and family (*16*). This makes social media a viable data source for user-generated content, particularly during a disaster when traditional surveys are inconvenient. Many literatures have demonstrated the effectiveness of social media data in disaster management, particularly in crisis communication (*16,24-29*), human mobility (*30-33*), damage valuation (*34*), and event detection (*35,36*), among others. However, there are very few studies that have looked at the social network connectivity of the local people and network-level properties of the local communities along with their crisis narrative analysis during a compound hazard like hurricane Laura amid the Covid 19 pandemic.

Twitter recently launched its Academic Application Programming Interface (API) (*37*), which provides a full history of public conversation through full-archive search endpoint (*38*), thus making Twitter a reliable social media data source. In this study, a user-mention network has been developed from the local Twitter user who tweeted during hurricane Laura and tweeted from the affected areas. The study is concerned with the following three research questions:

(i)     *Who are the agents of the social network, and what are their communication patterns?*
(ii)    *How does crisis communication differ from one community to another community?*
(iii)   *What is the role of different agencies in the social network?*

The findings of this study provide novel insights on how to create effective social media communication guidelines so that future disasters can be handled more effectively. The communication pattern in the local community is observed in this study with the local connectivity of different agents.

## RELATED WORK

According to global data portal by Kepois, the emergence of SMPs has been met by a spike in social media users of an excess of 4.70 billion people as of 2022, equating to ~59% of the total global population (*18*). People are becoming accustomed to relying more on social media than on mainstream news sources (*39*). SMPs act as a global connecting point for immediate response and news dissemination; for example, according to a recent study, Twitter alone generates more than 143K trending tweets per second (*40*). These SMPs are extremely effective communication tools, particularly during emergencies such as hurricanes, and they serve as a hub for public opinion and emotional guidance during crisis situations (*41*). Many researchers have used social media data in analyzing a variety of natural and man-made hazards, including fire (*42*), flood (*43*), tsunami (*44*), earthquake (*44,45*), hurricanes (*46,47*), and active shooting (*39,48*) among others.

Social media play a great role during an emergency like hurricane evacuation. The routing decision, evacuation destination, and evacuation decisions can be influenced by social media (*49*). Social media, as a new data source, can provide real-time information for flood monitoring (*43*). Extreme weather consequences can be seen in real-time on social media (*50*). Mohanty et al. found that the hurricane's per-capita economic impact is significantly correlated with per-capita Twitter engagement (*27*). In the classical evacuation literature, there have been a lot of studies on disaster response, particularly on hurricane evacuation (*51-59*) and wildfire (*60-62*), using traditional survey data, but according to recent studies, social media plays an important role in catastrophes by providing a participative, collaborative, and self-organized structure for communicating and disseminating situational information (*25,63,64*). The social network also has a great influence on decision-making, and decision-makers follow their peers during similar events (*65*).

Topic modeling, specifically Latent Dirichlet Allocation (LDA), is used in disaster-related information retrieval from social media data. Several studies used Twitter data to identify disaster-related topics using a static (*40,66-68*) and dynamic (*69*) LDA model. Yang et al. proposed a method for incorporating sentiment with a topic model (*70*); however, none of these studies consider any social network properties. Rajput et al. have explored the network properties of the different local agencies before, during, and after Hurricane Harvey (*71*). Sadri et al. found that users at the network's core are less eccentric and have higher degrees, and they are more engaged in disseminating information (*47*). It is pivotal to study how people interact in online social networks in order to develop and implement more effective guidelines in times of crisis (*67,72*).

However, very few studies have explored the connectivity of local people on social media during a compound crisis (a hurricane in the midst of a pandemic) as well as their communication patterns from a social network perspective. This study has explored the communication patterns of the locally affected people and connectivity with themselves and with different types of agencies from a user-mention network.

## DATA DESCRIPTION AND PREPROCESSING

There were several hurricanes in the year 2020. Hurricane Laura was the worst of them for the United States, killing at least ten people in Louisiana (*73*). On August 20, 2020, a tropical depression formed in the Cape Verde islands in the central Atlantic Ocean (*74*) and was officially named "Laura" on August 21, 2020 (*75*), which eventually became a category four hurricane and made landfall in Cameron, Louisiana, USA on August 27, 2020 (*76*). People freely express their opinion on social media like Twitter. Following is one of the tweets generated from Louisiana during the hurricane Laura: "*Me waiting patiently to get the internet and cable fix and back on.... I hate hurricane season*". One of the news channels tweeted regarding the devastation in the following tweet: "*@fox4beaumont Hurricane Laura death toll in Louisiana rises to 10*". In this study, Twitter data was used to identify crisis response communities in online social networks.

The Twitter Academic Application Programming Interface (API) (*37*), which provides a full history of public conversation through full-archive search endpoint (*38*), was used to get Twitter data from August 13, 2020, to September 03, 2020. Different query options have been used to make sure that all location-based data is collected. At first, '(Hurricane OR Hurricane Laura) has:geo' query option was used to collect all of the tweets regarding Hurricane Laura. Secondly, the "place: 'Louisiana'" query option was used to collect all of the generated tweets tagged with Louisiana within the mentioned time frame. Finally, the "point radius" query option with a 25-mile radius circle was used to ensure that all location-based tweets were collected from the highly impacted locations in Louisiana (**Figure 2**). The API sets several constraints for geolocation-based data collection, i.e., the maximum radius of the "point radius" cannot be greater than 25 miles (*38*). To overcome this constraint, multiple "point radius" queries were used on these highly affected locations.

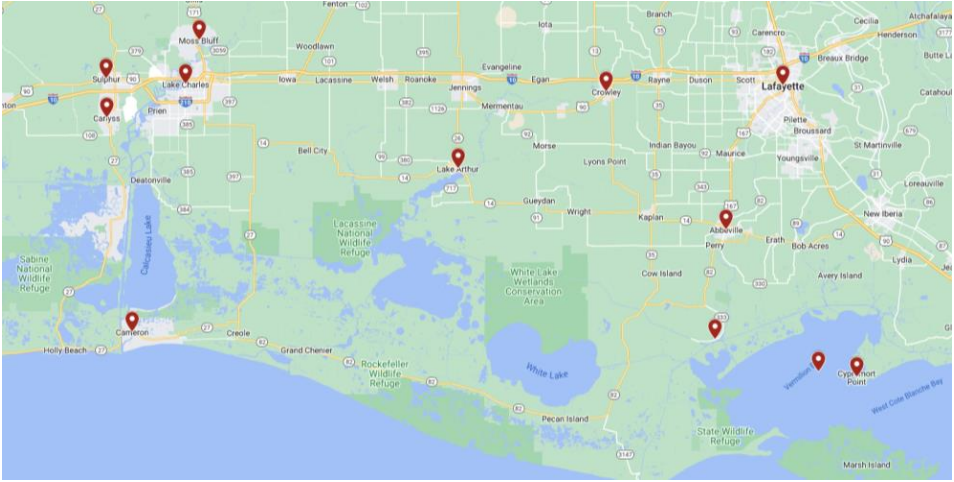

**Figure 2: Severely Impacted Areas in Louisiana by Hurricane Laura**

In this way, 1.3M tweets from around ~40k unique users have been collected. Since the study considered both location and keyword-based data collection approaches, Python's Pandas library was used to check for repetitions and ensure that all tweets in the data set were unique. To address the goal of this research, the study considered tweets created from the highly affected areas only, which in this case is only the Louisiana state. There are approximately 0.9 million tweets in the raw data where all of the tweets are geotagged. The language of all of the collected tweets was English. Elegant tweet preprocessing "*tweet-preprocessor*" Python package was used to remove noises: character codes, emojis, stop words, and html tags, etc. Then, all of the tweets were tokenized (broken down into smaller units: individual words or phrases). The study considered tweets relevant to hurricane Laura only. The relevance of a tweet was determined by identifying tokens within the tweet; more information on the steps and importance of relevance filtering can be found here (*77*). After all of this sorting and data preprocessing, the study finally got 0.4M tweets from ~20k unique users. The framework of this study is illustrated in **Figure 3**.

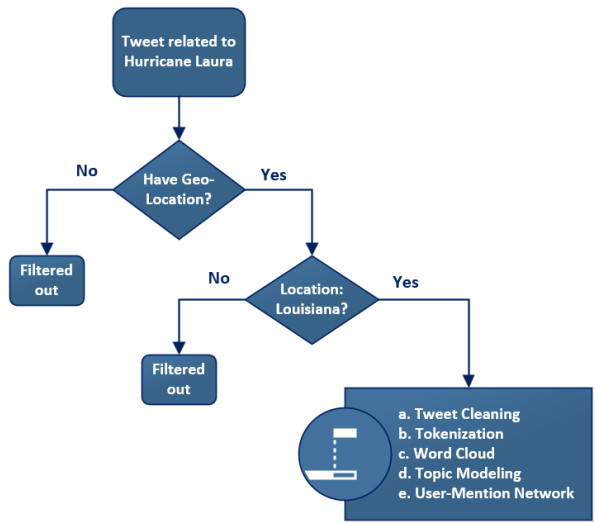

**Figure 3: Framework of the Study**

Every collected tweet has a bounding box location. At first, the centroid of the bounding box is determined, then using the reversed geo-code, these tweets have been mapped with the 2020 census block level county data. Out of the 64 counties in Louisiana, the study got tweets from almost all of them. The following **Figure 4** shows the county-level spatial distribution of the collected tweets.

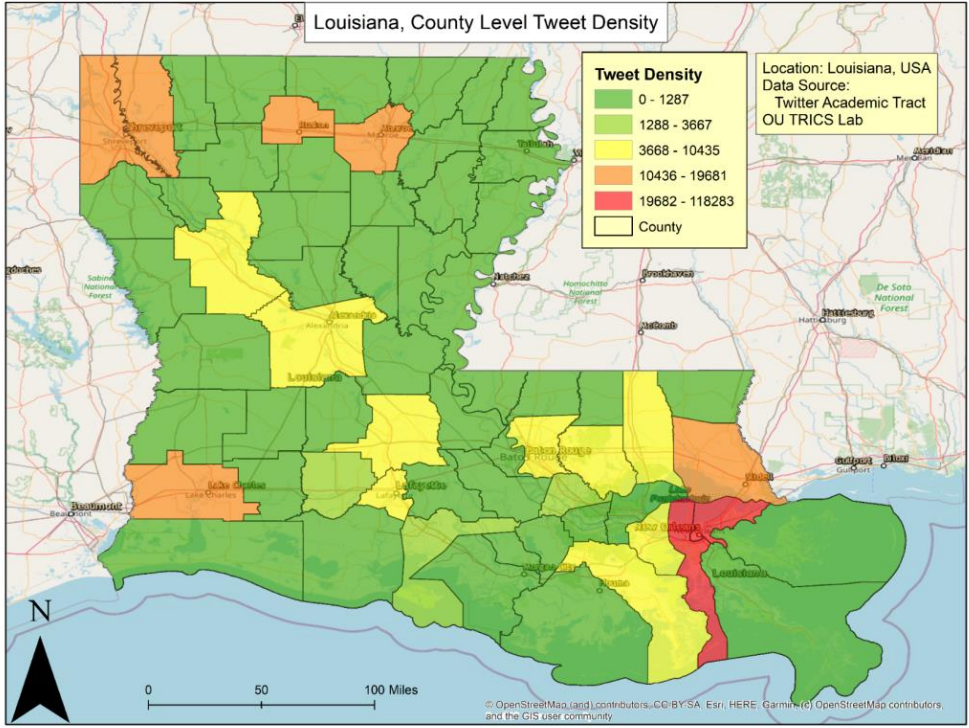

**Figure 4: County Level Tweet Density**

## METHODOLOGY

### Temporal Analysis with Sentiment Ratings

Twitter allows users to share the location of their tweets; if the device's location service is enabled, a confined area known as a bounding box is automatically generated. Tweet geolocation and timestamps were extracted using the *'geo'* and *'created at'* fields. Temporal or time series analysis is one of the most effective methods for comprehending data trends and temporal variation. Sentiment analysis, also referred to as opinion mining, is defined as the task of finding the opinions of authors of texts about specific entities (*78*). Sentiment analysis is a form of natural language processing (NLP) that involves emotion detection into three classifications: negative, neutral and positive that exist within value ranges -1, 0, and +1, respectively (*78,79*). This study uses the *VaderSentiment* analysis tool that is based on the Lexicon approach of sentiment analysis (*80*): more information regarding sentiment analysis can be found here (*81*). Word frequency, a popular NLP technique, counts how often a word appears in a text and orders the counts. Temporal heatmap is a sophisticated pivoting system that shows the term frequency over time. A color bar chart is used to represent frequency. The variations in color sequencing aid in distinguishing the frequency of the words.

### Word Bigram Analysis

Word bigram is an NLP concept that examines the likelihood of a duo of words appearing subsequent in a series of texts (*82*). It is capable of chunking word pairs and predicting the next forum word. The analysis separates the words arrangement *word₁...wordₙ* from the sequence of text observations $x_1...x_n$ for which the subsequent probability P(*word₁...wordₙ* | $x_1...x_n$) reaches its maximum. Equation 1 represents the algorithm.

$$arg\ max\ \{P(word_1 \dots word_N \mid x_1 \dots x_n) \cdot P(x_1 \dots x_n \mid word_1 \dots word_N)\}, word_1 \dots word_N \tag{1}$$

here, $P(x_1...x_n \mid word_1...word_N)$ denotes the conditional-probability of given the word arrangement $word_1...word_N$. Using the conditional probabilities, we obtain the decomposition as Equation 2

$$P(word_1 ... word_N) = \prod_{n=1}^{N} P(word_n \mid word_1 ... word_{n-1}) \tag{2}$$

A vocabulary of size W can be divided into $G_c$ word classes. $G_c$: $w_0 \rightarrow G_w$ represents the category mapping. Every word "w" in the vocabulary is associated with the word class $G_{w0}$. We use ($G_{v0}$, $G_{w0}$) to denote the equivalent class bigram for a word bigram ($v_0$, $w_0$). The equation for maximum likelihood estimate could be expressed as Equation 3

$$F_{bi} (\text{ß}) = \sum_{w} N(w) \, log \, N(w) + \sum_{G_v, G_w} log \, \frac{N(G_v, G_w)}{N(G_v)N(G_w)} \tag{3}$$

Given, $F_{bi}$ = bigram maximum likelihood; $N$ = training corpus size; w = size of vocabulary; *(u, v, w, x)* words in a document; ß = number of word classes *(82)*.

**Topic Modelling**

    The topic model framework in NLP examines empirical themes or subjects in text-based data. David Blei explained this using LDA *(39)*. It is based on the assumption that the collection of texts exposes a wide range of subjects. A topic model is used to find hidden semantic structures in text bodies. It clusters similar phrases into themes or topics. Based on word statistics in each topic, LDA establishes possible themes and balances the subjects of each document *(83,84)*. Griffiths and Steyvers proposed using Gibbs collapsed sampling to estimate the posterior distribution of word-to-subject or theme assignments P(z|w) *(85)*. The derivation steps for the update function for a new topic assignment of a word in sampling are summarized in the following equation.

$$P(z = t|z, w, \alpha, \beta) \propto \frac{n(d, t) + \alpha}{n(d, t) + T\alpha} \frac{n(t, w) + \beta}{n(t, w) + W\beta} \tag{4}$$

The number of assignments of word "*w*" in topic "*t*" is "*n(t, w)*", and the number of assignments of topic "*t*" in document "*d*" is "*n(d, t)*"; all counts exclude the current assignment "z". The descriptions of the various variables are as follows:

        n = count of tokens
        t = count of topics
        d = count of documents
        *w* = count of distinctive words
        $\theta$ = d × t of probabilities; topic distribution in documents
        $\varphi$ = t × *w* of probabilities; word distribution in topics
        $\alpha$ = d × t of $\alpha$ priors; Dirichlet prior for $\theta$
        $\beta$ = t × w of $\beta$ priors; Dirichlet prior for $\varphi$
        w = n-vector of word identity w; words in documents
        z = n-vector of topic assignment z; topic assignment of words

**Graph Theory**
*Component and Community*
    In graph theory, a connected subgraph that is not a part of any other bigger connected subgraph is referred to as a component of an undirected graph; for example, the network shown in **Figure 5 (a)** has three components. On the other hand, a community is a subset of nodes that are closely connected to one another and distantly connected to the nodes in the other communities. The network shown in **Figure 5 (b)** has two communities.

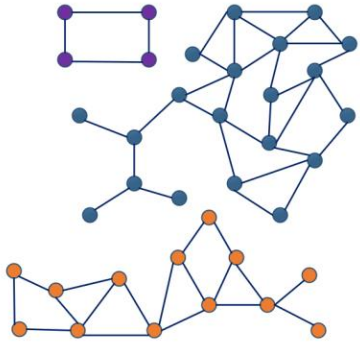

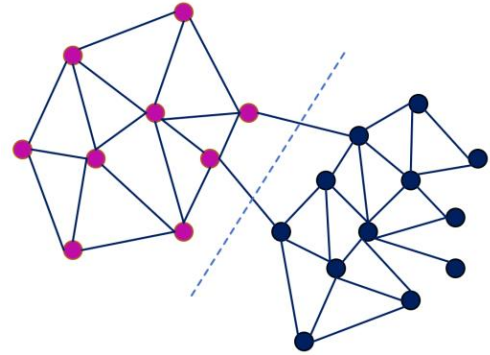

**Figure 5 (*a*): Components in Network**     **Figure 5 (b): Community in Network**

### Community Detection Algorithm:

Communities are a feature of many networks in which a single network may have numerous communities, each with strongly coupled nodes. When examining various networks, it may be necessary to look for communities inside them. Social media algorithms can utilize community detection techniques to find people who share common interests and keep them connected. Agglomerative methods and Divisive methods are the two main types of community detection methods. Edges are added one by one to a network that only comprises nodes in Agglomerative methods. From the stronger edge to the weaker edge, edges are added. Agglomerative procedures are followed by dividing methods. There, edges from a whole graph are eliminated one by one. In a particular network, there can be any number of communities, and they can be of various sizes. These qualities make community detection extremely difficult (*86*).

For each pair i,j of vertices in the network, a weight $W_{ij}$ is calculated initially, which denotes how tightly the vertices are related in some way. Because there are an infinite number of paths between any two vertices (unless it is 0), paths of length *l* are often weighted by a factor $a^l$ with *a* small, causing the weighted count of the number of paths to converge. If A is the network's adjacency matrix, and $A_{ij}$ is 1 if an edge exists between vertices i and j and 0 otherwise, then the weights, W in this formulation are determined by the matrix's elements.

$$W = \sum (aA)^l = [I - aA]^{-1} \qquad (5)$$

### Subgraph Generation

There are various types of networks based on the type of social media activity. Followee-Follower and user-mention network are the two most popular social networks (87). The followee-follower network is an unweighted directional network, whereas the user-mention network is a weighted directional network. Here, the weightage is determined by how many times a particular user has been mentioned. Information is passed from the user to the followees in a followee-follower network, and information is passed from the user to the mentioned person in the user-mention network. A user-mention network is more engaging than a followee-follower network because in a user-mention network, a specific user is directly mentioned, and the specific information is relevant to the user, whereas in a followee-follower network, there is a chance of missing any post or status of the person following, and the information might not be relevant for all of the followers. Based on the largest component, a subgraph was created from the main user-mention network in this study. The study examined various structural properties of the subgraph using network science concepts.

## RESULTS
### Temporal heatmaps with sentiment type

**Figure 6** shows the temporal heatmaps of tweet categories: positive, negative, or neutral. The heatmap shows that the study received at least 500 tweets per day from each category. At first, the number of neutral tweets was high, but as time passed, the number of negative tweets increased, and then after

Hurricane Laura made landfall on August 27, 2020, negative tweets peaked and then declined. Until August 19, 2020, the number of positive, negative, and neutral tweets was very close, but there has been a substantial increase in negative tweets, creating a significant difference between positive and negative tweets. The cause of such an increase could be hurricane Laura's devastation: property damage, utility disruption, high storm surge, and people's fear. After August 29, 2020, the number of positive and neutral tweets increased in comparison to negative tweets, possibly because the recovery phase began, and people received support and relief from local agencies.

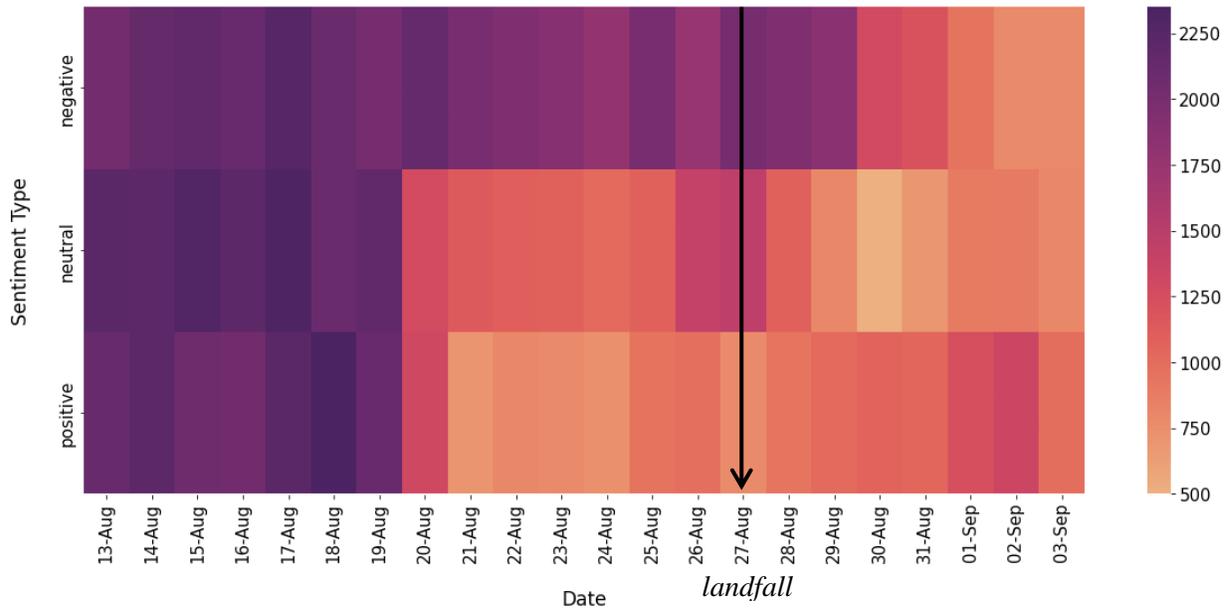

**Figure 6: Temporal heatmaps of tweet sentiment type**

**Temporal heatmaps of frequently used words**

The most frequently used words in a pool are sorted using frequent word analysis. The buzzwords in the texts tell the story. Heatmapping frequent words show how they are stranded over time. It shows how words rise and fall over time. For the top hundred most frequently occurring words (which is nearly ~46.45% of the total words), temporal heatmaps have been created and illustrated in two figures (**Figure 7** and **Figure 8**). The words were deemed in alphabetical order; they act as a contextual "dictionary," incorporating all aspects of conversation that may arise in discussions. The word frequency scale ranges from the most frequently occurring word "hurricane" to the 100th most frequently occurring word "work". The top five buzzwords are "hurricane", "emergency", "pandemic", "food", and "covid". The top frequent words of the heatmap reveal some obvious findings with the keywords: it is no coincidence that three of the top five buzzwords are related to the Covid 19, indicating a compound hazard situation created by hurricane Laura's landfall in the midst of the Covid 19 pandemic.

On the other hand, the sequence in which they show up in the heatmap provides an intriguing picture of the data. In the first heatmap (**Figure 7**) - three distinct patterns are clearly noticeable over three different time periods: before the landfall, days close to landfall, and after the landfall. There are a few words like "covid," "emergency," "hurricane," "pandemic," and "tornado" that were constantly repeated during the whole period of hurricane Laura. But there were a few words, like "power", "damage," "roof", "center," etc., that were not used as much in the beginning, but as the date of the landfall approached, their frequency drastically increased. It acts as evidence of the devastation hurricane Laura caused. When Hurricane Laura made landfall, people were not initially concerned about power outages, property damage, and utility disruption, but after the hurricane had made landfall, they began discussing it. Few words like "storm", "warning", "water", "wave", "weather", "wind," etc., were mentioned a lot before the landfall on August 27, 2020, but were less seen after the landfall. Most of the words in the second heatmap (**Figure 8**) were about the weather, news update, community, and properties.

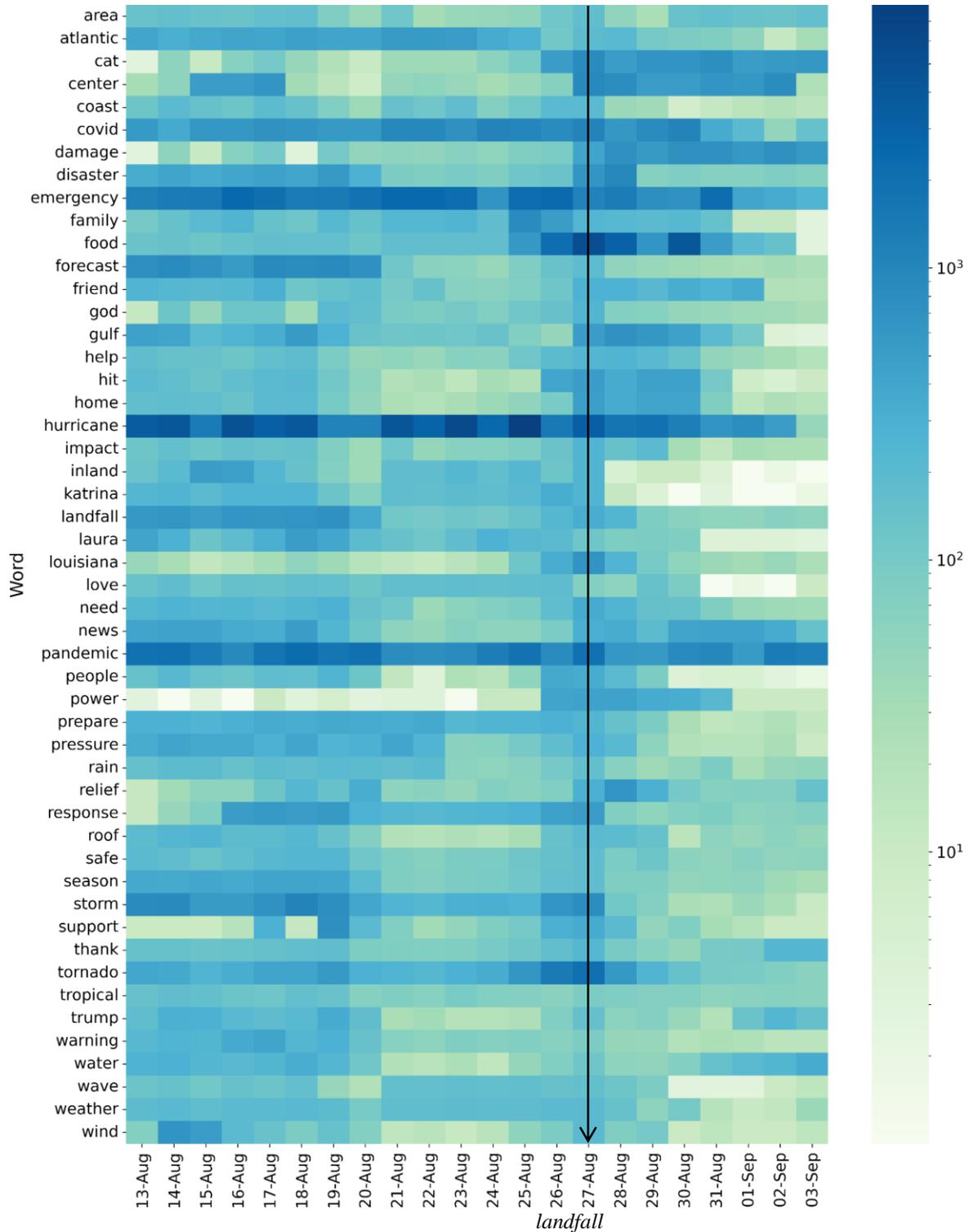

**Figure 7:** **Temporal heatmaps for top 50 words**

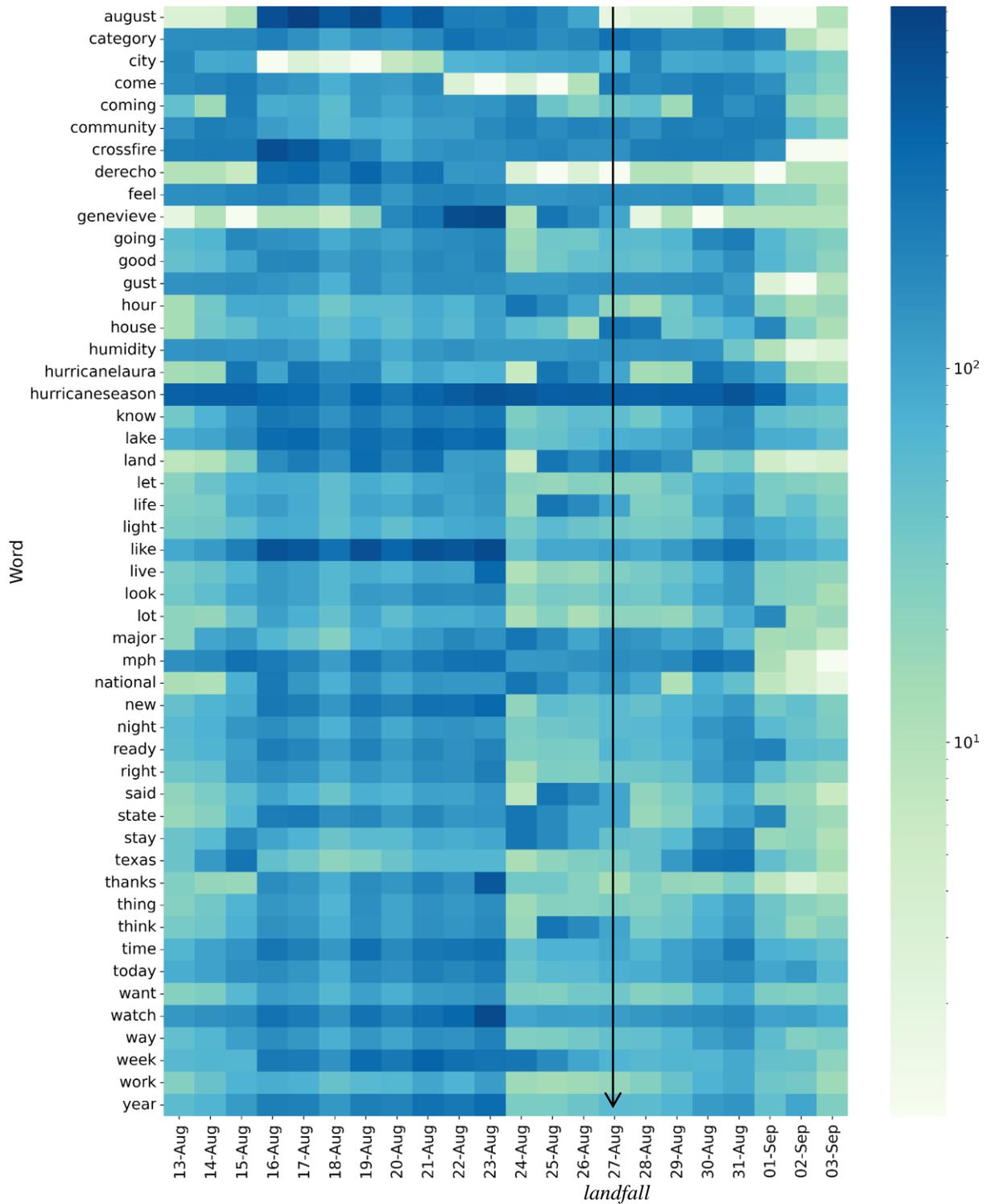

**Figure 8: Temporal heatmaps for top 51 to 100 words**

**User-Mention Directed Network**

The user-mention directed network is illustrated in **Figure 9**. The graph has 10,614 nodes and 16,923 edges. The degree of a node in a network indicates the connectivity: a higher degree means it is connected to more other nodes. The average degree of the network is $1.69 \cong 2$, which means that, on average, out of 10,614 people, each person is connected to nearly two others. This indicates that the majority of nodes are isolated and have formed small scattered groups. A network's diameter indicates the shortest path between two distant points. The diameter of this network is 12, which means it will take 12 steps to get from one point to the farthest point, which is significantly large than the six degrees of separation theory (*88*). The network consists of 78 communities and 21 components. The label of **Figure 9** depicts the component number and the number of nodes within that component. The color of the network represents the communities. There are 78 distinct colors in the network.

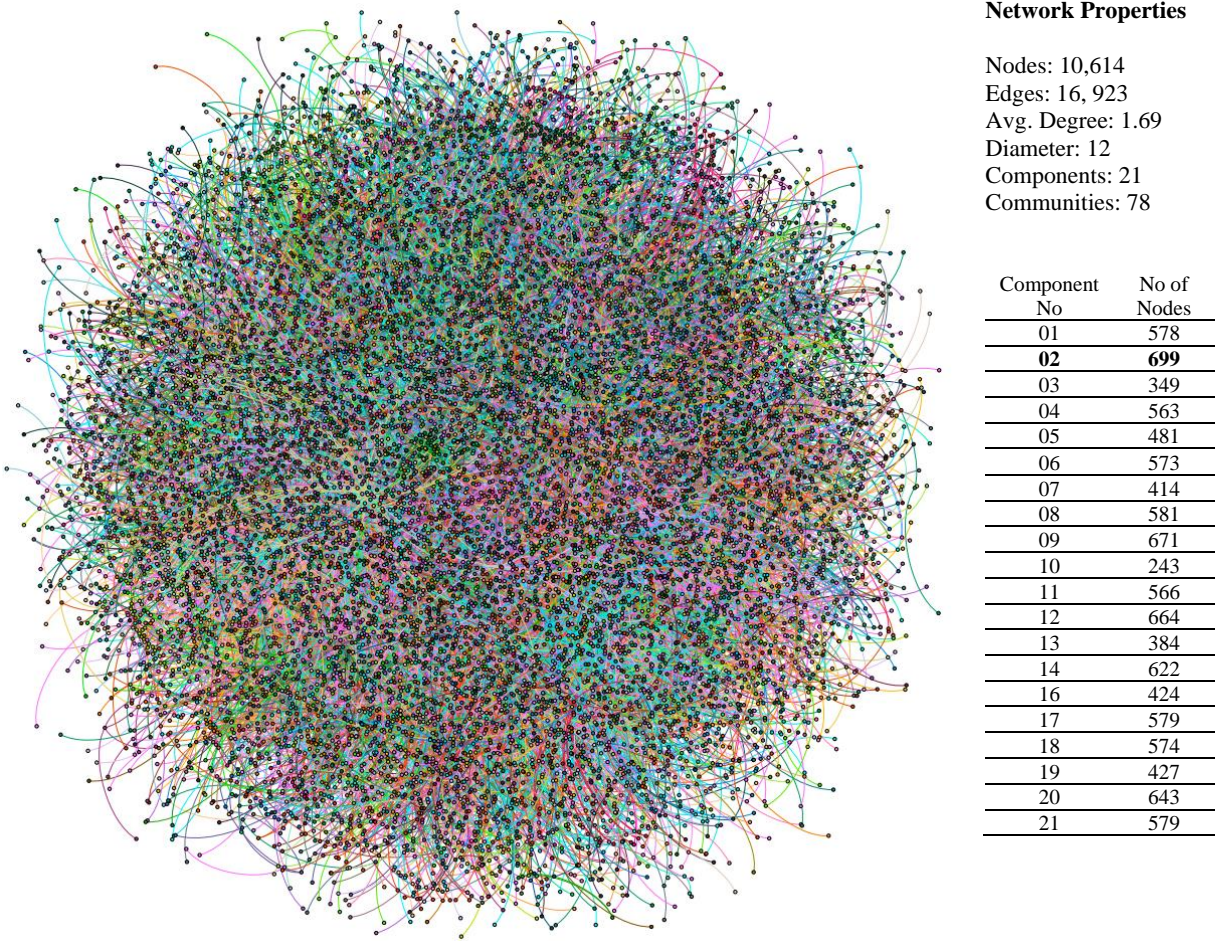

**Network Properties**

Nodes: 10,614
Edges: 16, 923
Avg. Degree: 1.69
Diameter: 12
Components: 21
Communities: 78

| Component No | No of Nodes |
|---|---|
| 01 | 578 |
| **02** | **699** |
| 03 | 349 |
| 04 | 563 |
| 05 | 481 |
| 06 | 573 |
| 07 | 414 |
| 08 | 581 |
| 09 | 671 |
| 10 | 243 |
| 11 | 566 |
| 12 | 664 |
| 13 | 384 |
| 14 | 622 |
| 16 | 424 |
| 17 | 579 |
| 18 | 574 |
| 19 | 427 |
| 20 | 643 |
| 21 | 579 |

**Figure 9: Hurricane Laura User-Mention Directed Network**

Component 2 has the most nodes, 699, with 578 unique users, while component 10 is the smallest one, with 243 nodes only. This study focused on the largest component for contextual analysis by generating a subgraph from the main user-mention network; however, this study can be scaled up or down depending on the requirements. **Figure 10** depicts the largest components of the network. The community determines the network's distinct hue. There are a total of eight communities. The network's node size is proportional to its degree; the higher the degree, the larger the node size. Community 6 has the largest network in terms of the total edges (730) in comparison to other communities. There is ~21% population alone in community 6, and community 1 is the smallest one with only ~3.5% population. The study also categorized the individuals in this subgraph according to their type, such as incident management, food bank, educator, journalist, online news portal, social organization, television news channel, etc. The study also categorized nodes into sixteen distinct agency types. The top five individuals from each community have been labeled in the network.

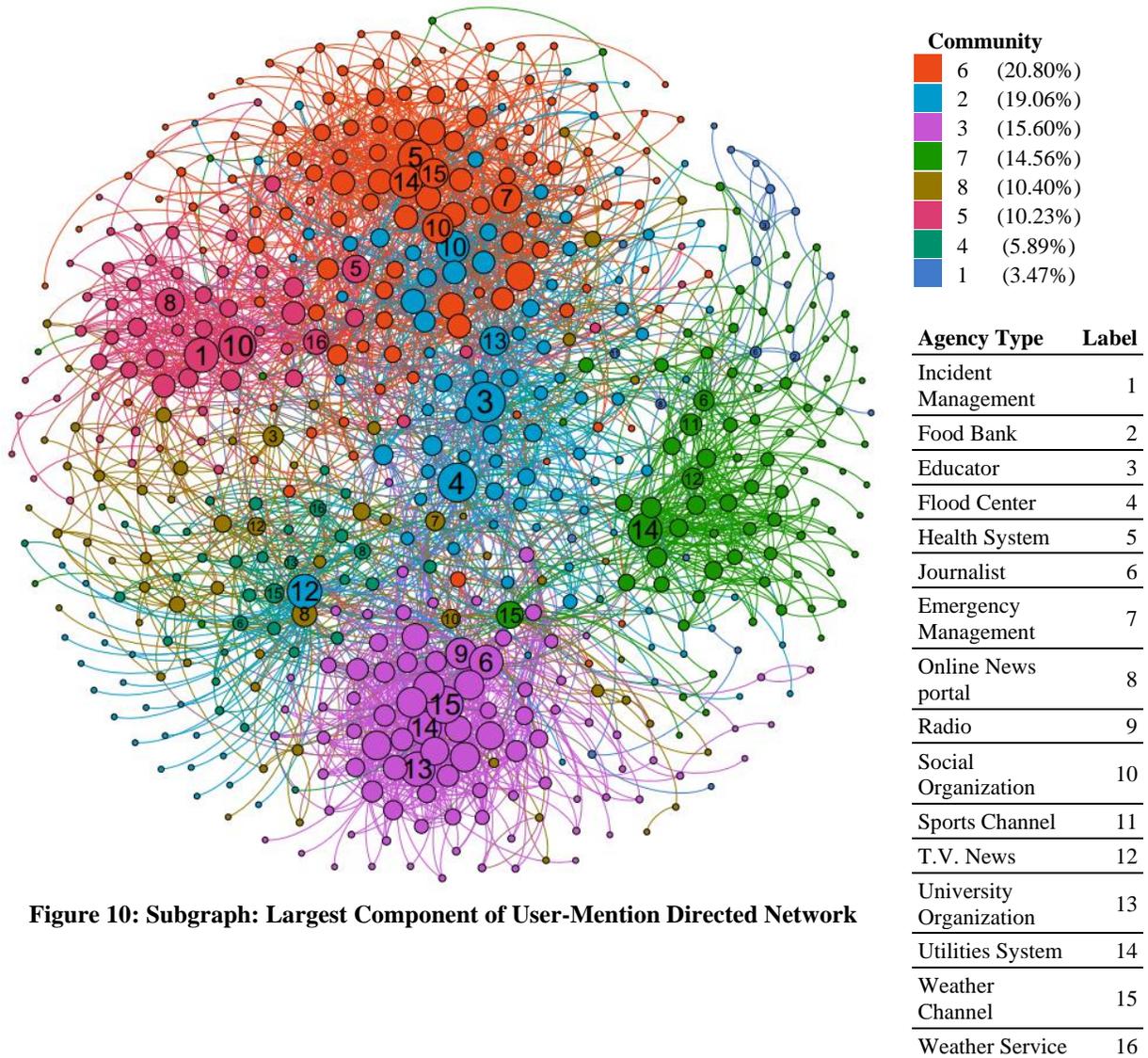

| Community | |
|---|---|
| 6 | (20.80%) |
| 2 | (19.06%) |
| 3 | (15.60%) |
| 7 | (14.56%) |
| 8 | (10.40%) |
| 5 | (10.23%) |
| 4 | (5.89%) |
| 1 | (3.47%) |

| Agency Type | Label |
|---|---|
| Incident Management | 1 |
| Food Bank | 2 |
| Educator | 3 |
| Flood Center | 4 |
| Health System | 5 |
| Journalist | 6 |
| Emergency Management | 7 |
| Online News portal | 8 |
| Radio | 9 |
| Social Organization | 10 |
| Sports Channel | 11 |
| T.V. News | 12 |
| University Organization | 13 |
| Utilities System | 14 |
| Weather Channel | 15 |
| Weather Service | 16 |

**Figure 10: Subgraph: Largest Component of User-Mention Directed Network**

It is evident from the figure that in between some nodes, there are two edges. It is due to the fact that they mentioned each other in their conversation. The single edge between two nodes indicates that the first node has only mentioned the second node, but the second node didn't mention the first one. The study also observed each community separately in order to better understand their concern and communication pattern. **Figure 11** illustrates the eight communities that comprise the subgraph network. Identifying the agents of the social network is one of the research questions of this study. These are the prominent agents in each of the following communities.

Community 1:   Sports page (L.C. Pride Basketball), Athletes, online news portal (12 News now), journalist, Food Bank to Fight Hunger

Community 2:   T.V. news channel (CNN), Educator, Louisiana watershed flood center, social organization (Kappa Sigma)

Community 3:   National Weather Service Office Lake Charles, University of Louisiana at Lafayette Black Male Leadership Association, Radio Station (Mustang 107.1, 1063 radio Lafayette), KLFY-TV, Fox 15

Community 4:    National Incident Management Systems and Advanced Technologies, News, Beacon Club

Community 5:    Louisiana Business Emergency Operations Center, The City-Parish of Lafayette, KPLC7News

Community 6:    Memorial Health System, Lafayette Utilities System, KATC TV, Latest News

Community 7:    KLFY-TV, Utility System, Townsquare Media Station, McNeese Football

Community 8:    News @ Lake Charles, McNeese State University, Fox 4, United Way of Acadiana

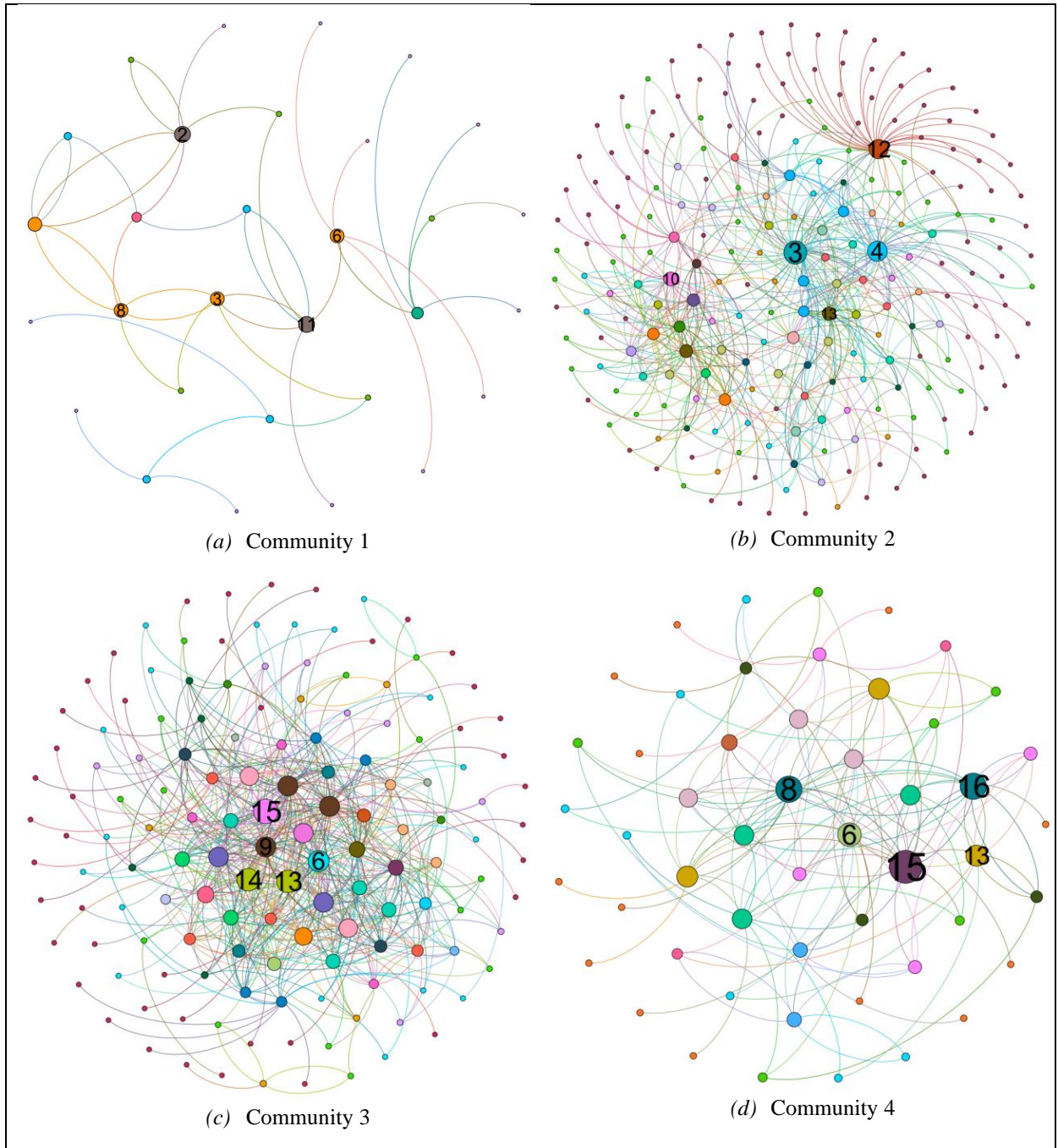

*(a)* Community 1                                   *(b)* Community 2

*(c)* Community 3                                   *(d)* Community 4

**Figure 11 (a): Communities of the Network (Community 1 to 4)**

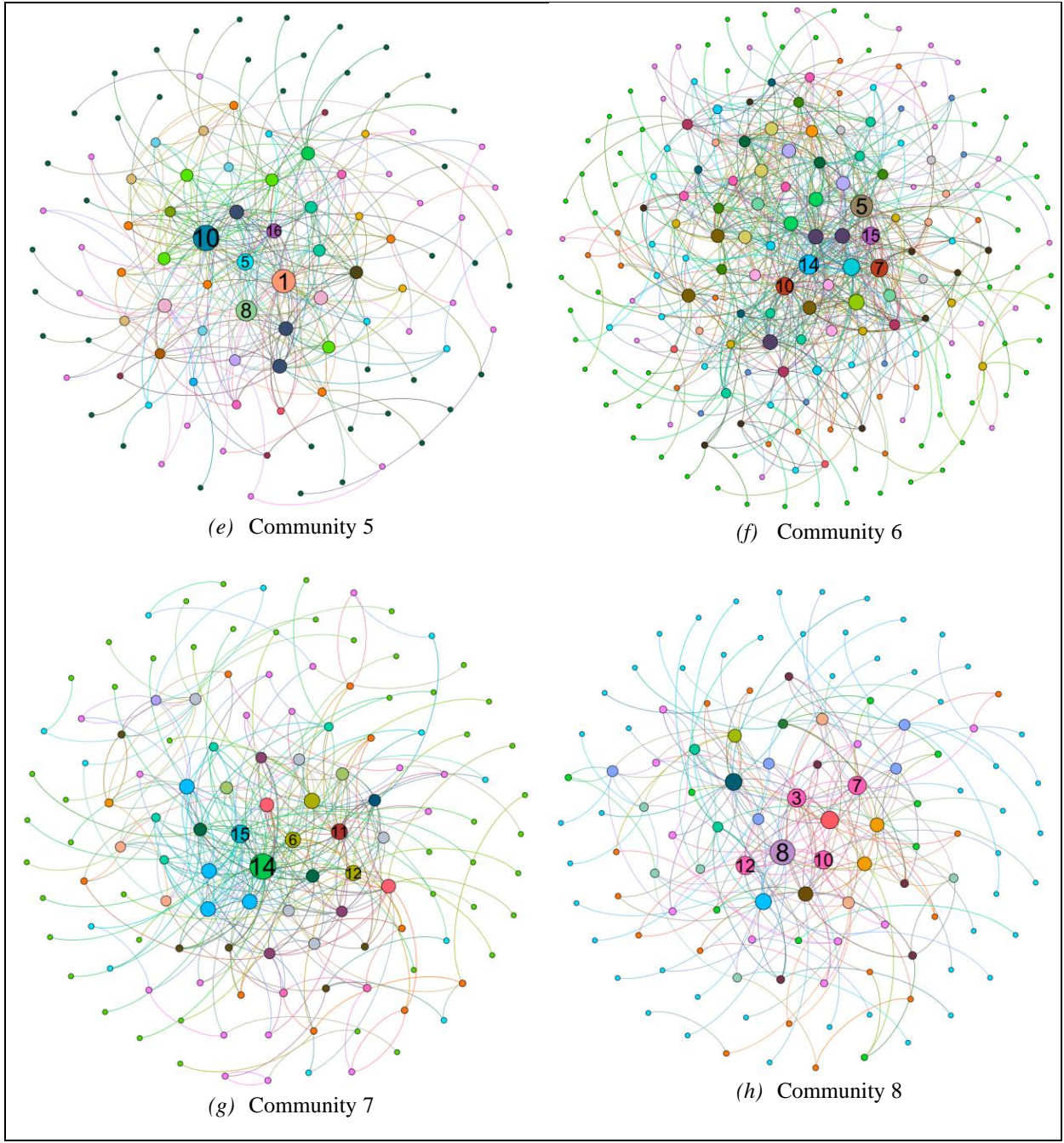

*(e)* Community 5

*(f)* Community 6

*(g)* Community 7

*(h)* Community 8

**Figure 11 (b): Communities of the Network (Community 5 to 8)**

The smallest community, community 1, has 30 nodes and 41 edges, a network diameter of 7, and an average degree of 1.367. Community 2 has 227 nodes and 576 edges. Despite having the most nodes, community 2 has a graph density of only 0.011, with an average degree of 2.537 and a network diameter of 8. With 142 nodes and 667 edges, community 3 has the highest edge/node ratio of 4.69. Community 3 has the highest graph density of 0.033, trailing only community 4, which has 51 nodes and 140 edges and a graph density of 0.055. As a result, community 4 has the highest population density of all. Community 5 has 108 nodes and 359 edges, with a graph density of 0.031 and a network diameter of 7. With a graph density of 0.031 and network diameter of 7, community 5 has 108 nodes and 359 edges. Community 6 has the highest number of edges (730) and the second-most nodes (198), after community 2. The graph density of community 7 (126 nodes and 356 edges) and community 8 (116 nodes and 248 edges) is 0.023 and 0.019, respectively.

**Word Bigram Network Mapping**

A word bigram is a pair of words that are next to each other in a text. **Figure 12** shows which two words are most frequently appeared together in the corpus. The study used a network to map the bigram words to gain a better understanding of their conversation in each community. The degree determines the color of each node in the network. The centered word has the highest degree, for example, in community 1, the words *charles*, *Laura*, and *hurricane* have three degree, and the rest of the words, with the exception of *catastrophic*, *jesus*, *helplakecharles* and *days*, have two degree. These words depicted in the bigram are buzzwords in each community and tell what was being cooked in each community.

In community 1, the word "*power*" headed by *"without"* and succeeded by "*water*" indicates the utility disruption, and the phrases "*lake-city-destroyed-completely*", "*laura-entire-wiped-city*", and "*roof-destroyed*" refer to the damage that hurricane Laura did to Lake Charles in Louisiana. Hurricane Katrina, a Category 5 Atlantic hurricane that made landfall on August 29, 2005, was also a traumatic event for Louisiana. After 14 years, on August 27, 2020, they experienced something similar; the words "*august-issued*", and "*two-hurricanes*" in community 2 bring up this issue. The bigram network clearly shows that the primary topic of conversation in each community is as follows:

Community 1:   Utility disruption, property damage, flooding
Community 2:   Utility disruption, property damage, comparison with Hurricane Katrina
Community 3:   Evacuation, utility disruption, relief seeking, Covid-19 pandemic, property damage
Community 4:   Seeking donation (fundraising), utility disruption, property damage
Community 5:   Compound hazard, utility disruption, Cameron, weather, death
Community 6:   Election campaign, public property damage, insurance, relief, recovery
Community 7:   Media, reports, weather, property damage, help-seeking
Community 8:   Relief transportation, utility disruption, help-seeking, insurance, gas, return to home

The most common topic of discussion in each community is utility disruption (i.e., power, internet, water, gas) and property damage; however, each community has its own distinctive conversations also. A word bigram is a great tool for understanding the conversation's latent meaning; however, topic modeling is a more established method in natural language processing for understanding the inherent topics.

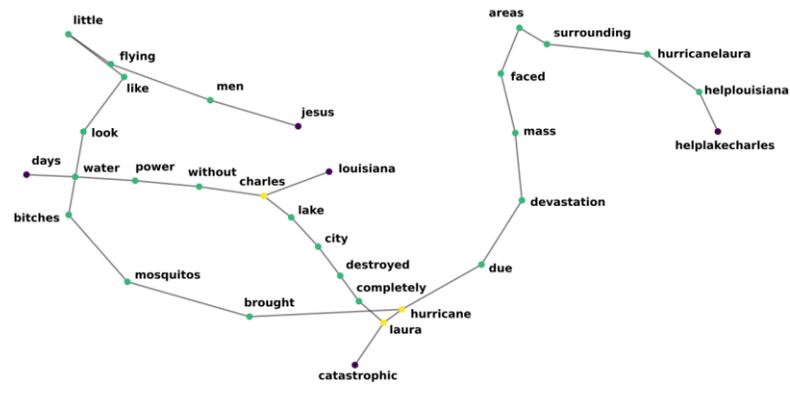

*Community 1*

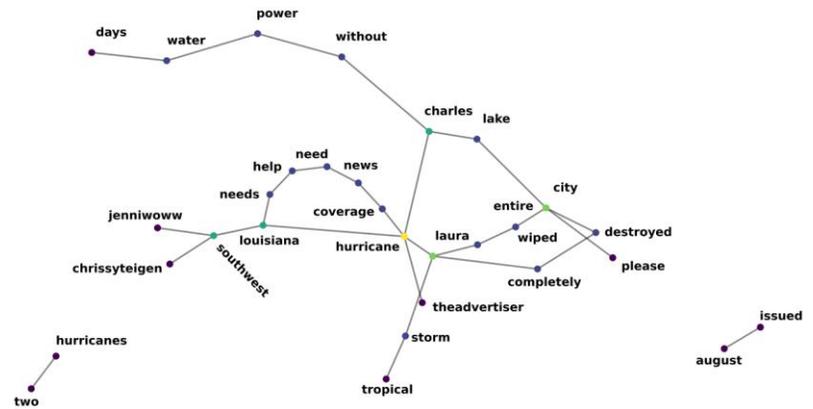

*Community 2*

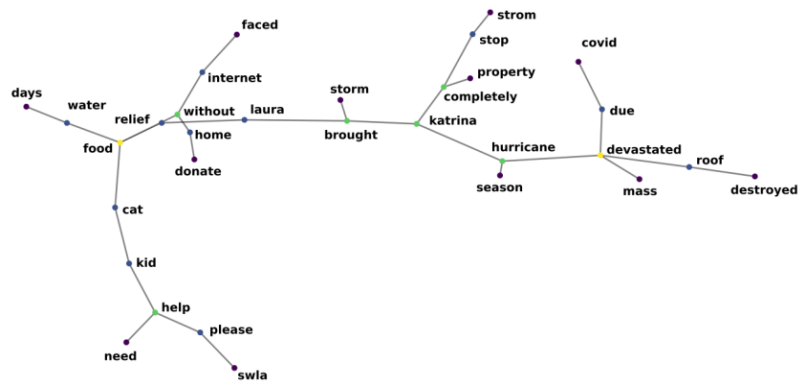

*Community 3*

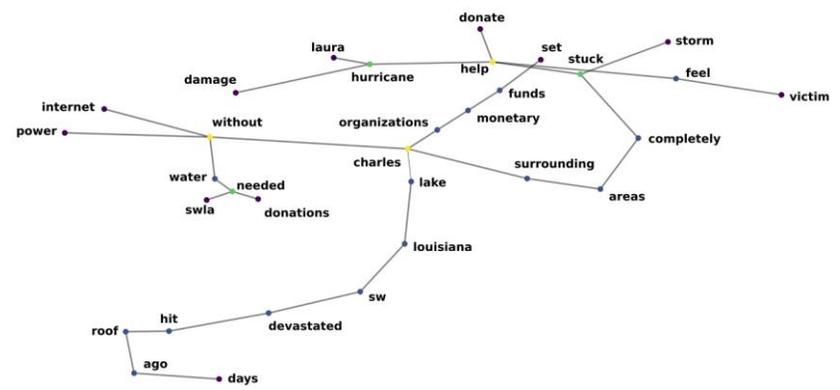

*Community 4*

**Figure 12 (a): Bigram Network: Community 1 to 4**

*Community 5*

*Community 6*

*Community 7*

*Community 8*

**Figure 12 (b): Bigram Network: Community 5 to 8**

# TOPIC MODELING

Topic modeling is a statistical model that uses a probabilistic process based on a set of sampling rules to generate abstracts of topics (*16*). This procedure describes how latent variables generate database topics. Perplexity, loglikelihood, coherence value, etc., can measure model performance. The study assessed the coherence value for topic model optimization. Each topic model has an overall coherence value, as well as a separate score for each topic within the model. The higher the coherence value, the better the model's performance. When the coherence value reaches its maximum, the topics begin to reoccur in the model, reducing its efficiency. Tweets in each community have been analyzed for topic modeling, and each topic generated ten salient words, of which the top five significant words with their probability are listed in **Table 1**.

**Table 1: Highlights of Topic Modeling**

| Community | | Probable Topic | Most Probable Words in Coherent Topic |
|---|---|---|---|
| 1 | Topic 1 | Utility interruption in Lake Charles | lake (0.356), charles (0.356), destroyed (0.115), power (0.096), completely (0.088) |
| | Topic 2 | Tropical storm in Louisiana | louisiana (0.122), storm (0.107), lafayette (0.090), tropical (0.087), going (0.075) |
| 2 | Topic 1 | Urging help | need (0.206), helplouisiana (0.116), devastation (0.063), surrounding (0.052), coverage (0.131) |
| | Topic 2 | Power interruption | power (0.214), without (0.189), completely (0.173), outage (0.094), day (0.053) |
| | Topic 3 | Aftermath of hurricane | parish (0.122), southwest (0.107), really (0.090), aftermath (0.087), damage (0.075) |
| 3 | Topic 1 | Relief for lake Charles | help (0.143), relief (0.132), hand (0.127), food (0.119), water (0.077) |
| | Topic 2 | Prior experience | katrina (0.141), completely (0.141), property (0.140), brought (0.133), ago (0.076) |
| | Topic 3 | Compound disaster | covid (0.719), pandemic (0.117), hurricane (0.089), laura (0.078), season (0.045) |
| 4 | Topic 1 | Fundraising | funds (0.205), monetary (0.128), organizations (0.108), donations (0.094), set (0.068) |
| | Topic 2 | Water captivity | stuck (0.462), water (0.362), surrounding (0.214), days (0.122), feel (0.079) |
| 5 | Topic 1 | Extent of destruction | hurricane (0.205), laura (0.127), landfall (0.108), category (0.094), killed (0.068) |
| | Topic 2 | First landfall | cameron (0.239), mph (0.111), wind (0.097), hurricane (0.091), week (0.089) |
| | Topic 3 | Compound hazard | covid (0.189), expected (0.105), pandemic (0.096), emergency (0.071), help (0.065) |
| | Topic 4 | Media coverage | news (0.129), media (0.102), coverage (0.089), areas (0.52), cameron (0.039) |
| 6 | Topic 1 | Recovery update | recovery (0.252), macro (0.239), update (0.131), messages (0.126), covered (0.056) |
| | Topic 2 | Election campaign | trump (0.726), president (0.726), jefferson (0.245), county (0.124), vote (0.089) |
| | Topic 3 | Weather forecast | wind (0.215), coverage (0.207), forecast (0.178), tropical (0.125), key (0.105) |
| | Topic 4 | Agency mention | fema (0.287), support (0.148), offering (0.114), wind (0.078), key (0.062) |
| | Topic 5 | Damage in School | school (0.180), roof (0.168), survived (0.163), southeast (0.056), high (0.031) |
| **7** | Topic 1 | Weather Report | weather (0.127), news (0.112), report (0.082), wind (0.068), rain (0.039) |
| | Topic 2 | Seeking help | help (0.125), really (0.103), could (0.095), devastation (0.063), surrounding (0.057) |
| | Topic 3 | Utility interruption | power (0.211), without (0.179), completely (0.145), outage (0.089), day (0.031) |
| | Topic 4 | Seeking help | need (0.213), helplouisiana (0.103), devastation (0.073), surrounding (0.061), coverage (0.034) |
| **8** | Topic 1 | Relief | truck (0.247), goods (0.237), lifted (0.126), relief (0.111), lousiana (0.056) |
| | Topic 2 | Home Coming | return (0.148), busy (0.126), home (0.078), needed (0.031), days (0.021) |
| | Topic 3 | Utility interruption | power (0.124), gas (0.095), without (0.078), insurance, (0.063) electricity (0.59) |

From the eight communities, a total of 26 topics were found. Community 1 has the fewest number of topics (2 topics), while community 6 has the most (5 topics). The majority of the communities talked about utility outages, tropical storms, property damage, seeking assistance, and rising funds.

## CONCLUSIONS

Hurricane Laura devasted Louisiana and nearby areas in the midst of a pandemic. The preparedness of local people and agencies had to cope up with the situation. There are very few studies that have explored the connectivity of the local people in social media during a compound crisis, along with their communication pattern, from a network perspective. This study investigated the communication patterns of locally affected people, as well as their connectivity with themselves and various types of agencies, using data-driven approaches. The tweets of the local users revealed a crisis communication pattern based on their community. The number of agencies and users connected to a community reveals the nature of their discussion. The study analyzed the collected tweets using multiple natural language processing: (**i**) the top 100 frequently used words grasp the most buzzed words in the deliberations, (**ii**) temporal heatmaps show how topics emerge and trend over time during and after the landfall, (**iii**) the directed user-mention network shows how local residents are connected to one another and to various agencies, (**iv**) community detection in the subgraph enables to uncover hidden associations among different network nodes, (**v**) word bigram network captures the centrality and correlation of the discussions and (**vi**) the topic modeling depicts the overall scenario of the compound hazard.

The crisis communication of the local people during hurricane Laura on Twitter shows how people interact and engage in social media during a major disaster and how they engage in connectivity with different agencies and organizations. The connectivity of the Twitter user was determined based on their mention in the original tweet. Some key findings of the study are summarized below:

- Based on the subgraph generated within the scope of the study, twenty-one network components of different sizes were observed in the subgraph network. The largest component included eight well-defined communities that revealed high levels of engagement between the public and a number of agencies/organizations. The communities discussed about utility outages, seeking assistance, compound disaster, water captivity, fundraising, and returning home.
- One of the communities that primarily comprised of health and utility systems along with some news media (KATC TV) was found to be the most connected, and community that comprised of T.V. news channel (KLFY-TV, Fox 15), National Weather Service Office Lake Charles, University Organization (University of Louisiana at Lafayette Black Male Leadership Association) and radio Station (Mustang 107.1, 1063 radio Lafayette) is the most populous network found in this study.

- Sports channels, university clubs, utility systems, weather channels, social organizations, health systems, and online news portals are the most connected (have the highest node degree) and influential node in their respective communities.
- Although emergency management and law enforcement agencies play a critical role in responding to such emergencies, the online communities found in this study revealed that university-related organizations, such as the University of Louisiana at Lafayette Black Male Leadership Association, Beacon club, social organizations such as Kappa Sigma, distinct journalist, T.V channel., online news portal, electronic media, weather service (National Weather Service Office Lake Charles), sports page such as L.C. Pride Basketball, McNeese Football, and some athletic were found to be very active in the network.
- The most common topic of conversation in each community is utility disruption (i.e., power, internet, water, gas), followed by property damage; however, each community has its own unique conversations also, such as evacuation, insurance coverage, election campaign, fundraising, and so on.

The twenty-one connected components found in this study indicate that there is twenty-one isolated group with poor engagement. In future crisis communications, this communication gap should be addressed. Again, the network diameter ranges from 7 to 9 in the subgraph's eight distinct communities. It indicates that transferring information from one actor to another will take at least seven steps, which exceeds the six-degree separation rule. The more steps required to transfer information from one actor to another, the greater the possibility of information deviation. In the future, the influential entities should be more engaged with each other and connect as many social media agents and/or individuals as possible. Emergency management and law enforcement agencies should be more active on social media, in addition to their physical presence. They can reach out to more people in a shorter period of time and learn about their necessities through social media.

The findings of this study provide novel insights that would allow developing effective social media communication guidelines so that future disasters can be handled more effectively. The network analysis in this study, however, was limited to the largest component, which can be expanded to investigate other components and communities of the network. In addition, this study considered all topics to be static, i.e., time-independent, which cannot explain the topic's evolution over time. Future research may utilize more advanced topic modeling techniques to capture the dynamics of word distribution within a topic.

## ACKNOWLEDGMENTS

This material is based upon work supported by the National Science Foundation under Grant No. IIS-2027360. However, the authors are solely responsible for the findings presented in this study. The analysis and results section is based on the limited Twitter dataset and the authors' opinion, which cannot be expanded to other datasets without detailed implementation of the proposed methods. Other agencies and entities should explore these findings based on their application/objectives before using these findings for any decision-making purpose. Any opinions, findings, and conclusions, or recommendations expressed in this material are those of the author(s) and do not necessarily reflect the views of the National Science Foundation.

## AUTHOR CONTRIBUTIONS

The authors confirm that all authors contributed to the following aspects of the paper: study conception and design, data collection; data analysis and interpretation; and draft manuscript preparation. All authors reviewed the findings and approved the final manuscript version.